**METHODOLOGY**

**Open Access**

# Methodological concerns about "concordance-statistic for benefit" as a measure of discrimination in predicting treatment benefit

Yuan Xia[1*] , Paul Gustafson[1] and Mohsen Sadatsafavi[2,3]

## Abstract

Prediction algorithms that quantify the expected benefit of a given treatment conditional on patient characteristics can critically inform medical decisions. Quantifying the performance of treatment benefit prediction algorithms is an active area of research. A recently proposed metric, the concordance statistic for benefit (*cfb*), evaluates the discriminative ability of a treatment benefit predictor by directly extending the concept of the concordance statistic from a risk model with a binary outcome to a model for treatment benefit. In this work, we scrutinize *cfb* on multiple fronts. Through numerical examples and theoretical developments, we show that *cfb* is not a proper scoring rule. We also show that it is sensitive to the unestimable correlation between counterfactual outcomes and to the definition of matched pairs. We argue that measures of statistical dispersion applied to predicted benefits do not suffer from these issues and can be an alternative metric for the discriminatory performance of treatment benefit predictors.

## Background

Precision medicine emphasizes optimizing medical care by individualizing treatment decisions based on each patient's unique characteristics. A better understanding of the heterogeneity of treatment effect is the foundation for formulating optimal treatment decisions [1]. Treatment benefit predictors, mathematical functions that predict the average treatment benefit conditional on individuals' characteristics, are important enablers of such individualized care [2].

Much of the methodological development in predictive analytics has been based on risk predictors, functions that return an estimate of the risk of an event given patient characteristics. Focusing instead on the prediction of treatment benefit presents a relatively new paradigm. As such, there is an increasing interest in evaluating the performance of treatment benefit predictors. For risk prediction, the performance of a predictor is often categorized into discrimination, calibration, and clinical utility (net benefit) [3]. Discrimination refers to the ability of the predictions to separate individuals with and without the outcome of interest. Calibration is about how close the predicted and the actual risks are. Net benefit evaluates the clinical utility of a risk prediction model by subtracting the harm from the false-positive classification from the benefit from the true positive classification. The treatment benefit paradigm can incorporate these concepts. For example, calibration plots [4], discriminatory performance measures [5, 6], and net benefit [7] have been extended to evaluate treatment benefit predictors. Maas et al. further extended other performance measures, such as E-statistic, cross-entropy, and Brier score, to treatment benefit predictors [8].

*Correspondence:
Yuan Xia
lily.yuanxia@stat.ubc.ca
[1] Department of Statistics, University of British Columbia, Vancouver, Canada
[2] Respiratory Evaluation Sciences Program, Faculty of Pharmaceutical Sciences, University of British Columbia, Vancouver, Canada
[3] Faculty of Medicine, University of British Columbia, Vancouver, Canada

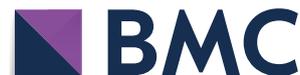

Our work focuses on metrics (i.e., summaries) for the discriminatory performance of treatment benefit predictors. In particular, the concordance statistic for benefit (*cfb*) proposed by van Klaveren et al. (2018) evaluates the discriminatory performance of treatment benefit predictors by conceptually extending the idea of the concordance statistic (c-statistic) for risk predictors [5]. The metric, *cfb*, has been used in a number of applied studies. For instance, Meid et al. (2021) used *cfb* to evaluate a treatment benefit predictor for oral anticoagulants for preventing strokes, major bleeding events, and a composite of both [9]. Duan et al. (2019) used *cfb* to compare two models for predicting individual treatment benefit for intensive blood pressure therapy [10].

In this work, we scrutinize *cfb* on multiple fronts. In particular, we consider its properness as a scoring rule, its sensitivity to the correlation between counterfactual outcomes, and the definition of matched patient pairs at the population level. In a recent preprint [11], Hoogland et al. also considered some theoretical and methodological issues around *cfb*. Some of their concerns connect with some of ours, as we will indicate later. The rest of this manuscript is structured as follows. We first review the original description of *cfb* [5] and provide an analogous definition at the population level. We demonstrate several scenarios in which *cfb* is shown to be an improper scoring rule, followed by further discussions.

## The concordance statistic for benefit (*cfb*)
### Notation
We consider scenarios arising from a binary treatment decision $T$ (control: 0 v. treated: 1) and a binary outcome $Y$ (unfavorable outcome: 0 v. favorable outcome: 1). The individual treatment benefit is usually formulated as the algebraic difference in outcomes under both treatment arms (i.e., treatment minus control). When the outcome is binary, individual treatment benefit can be described by a ternary variable $B$ with levels consisting of harm ($B = -1$), no effect ($B = 0$), and benefit ($B = 1$). In some narrow contexts, such as some cross-over studies where strong assumptions hold, $B$ can be directly observed. However, in most situations, $B$ is not observable. For instance, in a prototypical parallel-arm clinical trial, $B$ cannot be observed directly as one of the outcomes is counterfactual. van Klaveren et al. provided an algorithmic definition of *cfb* in such studies based on comparing outcomes from different treatment arms measured on two similar patients [5]. The definition of *cfb* in the above-mentioned work is based on a given sample. However, we stress that the descriptions of *cfb* involving phrases such as "the proportion of all possible pairs" and "probability that forms two randomly chosen matched pairs" make clear how *cfb* can be readily interpreted at the population level. In what follows, and without loss of generality, we view the *cfb* as a population-level attribute defined regardless of whether $B$ is observable or estimated.

A vector of baseline covariates for patients is denoted as $X$, and $E[B \mid X]$ is the average treatment benefit of a subpopulation stratified by $X$. Let a function of $X$ denoted by $h(x)$ be a treatment benefit predictor, i.e., $H = h(X)$ is taken as a prediction of $B$. The best possible predictor, denoted as $h^*(x) := E[B \mid X = x]$, is a special case in which the corresponding *cfb* is indicated as *cfb**. We restrict our attention to randomized controlled trials (RCTs) in which $H^* = h^*(X)$ can be expressed as

$$E[B \mid X] = E[Y \mid T = 1, X] - E[Y \mid T = 0, X],$$

ranging between $-1$ and 1.

### Definition of *cfb*
The original definition of *cfb* directly extends the definition of the c-statistic from risk predictors for binary outcomes to treatment benefit predictors [5]. To define *cfb*, we randomly select two patients from the population, whose treatment benefit quantities are $\{(B_1, H_1), (B_2, H_2)\}$. The pair is concordant if $(B_1 - B_2)(H_1 - H_2) > 0$. It is discordant if $(B_1 - B_2)(H_1 - H_2) < 0$. If $B_1 \neq B_2$ and $H_1 = H_2$, it is a tied pair. Otherwise, the pair is not considered. We score each concordant pair by 1, each tied pair by 0.5, and each discordant pair by 0. The *cfb* is the average of the scores. Mathematically, *cfb* can be expressed as

$$cfb = \Pr(H_1 > H_2 \mid B_1 > B_2) + 0.5 \Pr(H_1 = H_2 \mid B_1 > B_2). \quad (1)$$

In many applications ties among $H$ will not occur, for instance, if $X$ has a continuous component and $h(\cdot)$ is smooth. In this case, *cfb* is a proportion of concordant pairs over the pairs satisfying $B_1 \neq B_2$, which can be determined as $cfb = \Pr(H_1 > H_2 \mid B_1 > B_2)$. This was the working definition by van Klaveren et al. [5]. By incorporating ties, a technique commonly used in defining the c-statistic, the concept behind *cfb* can be extended, allowing for a more intuitive demonstration of its properties in various scenarios. When $H$ is independent of $B$, $cfb = 0.5$. Reciprocally, if a large proportion of pairs are concordant, indicating units receiving greater $B$ also have greater $H$, then the value of *cfb* is close to the maximum 1.

This definition of *cfb* is descriptive and directly reflects the concordance association between predictions and actual benefits. Its simplicity allows us to focus on the conceptual aspect of the *cfb*.

## Methodological concerns about *cfb*
### *cfb* is an improper scoring rule

The definition of the proper scoring rules dates back to Savage (1971) [12]. Gneiting and Raftery (2007) [13] defined the "proper scoring rule" for probability-type metrics, defining a metric as proper if the expectation of the metric is maximized (or minimized) when correct probabilities are used. This concept has since been frequently used to investigate the reliability of metrics [14, 15]. Pepe et al. (2015) expanded on the concept of properness for metrics that evaluate the improvement in prediction performance gained by adding extra covariates [16]. While the technical definitions of a proper scoring rule vary slightly across the works cited above, the spirit is the same. And the obvious adaptation to treatment benefit prediction context would be to demand that in any given population the summarizing metric, as a function of $h(\cdot)$, be maximized by $h^*(\cdot)$ that truly outperforms other benefit predictors (in the sense of expected squared error).

According to the provided definition of *cfb*, we start with a distribution of $(B, X)$. We consider a single binary variable $X$. As such, the distribution of $(B \mid X)$ can be summarized by two probability triples $\{(p_{-1}, p_0, p_{+1}), (q_{-1}, q_0, q_{+1})\}$ respectively for $X = 0$ and $X = 1$. Particularly,

$$p_i = \Pr(B = i \mid X = 0), \quad q_i = \Pr(B = i \mid X = 1),$$

where $i$ denotes the level of $B$, and $\sum_i p_i = \sum_i q_i = 1$.

An example might be breast cancer surgery, where the treatment is surgical therapy and the control is conservative therapy. The tumor grade $X$ is binary with values (0: low grade; 1: high grade), and the individual treatment benefit, $B$, is ternary. Assume $\Pr(X = 1) = 0.5$, and consider the distribution of $(B \mid X)$ summarized by two probability triples to be $\{(0.25, 0.01, 0.74), (0.14, 0.18, 0.68)\}$ with $E[B] = 0.515$ for the treatment.

Properties of *cfb* are investigated by comparing two treatment benefit predictors: $h^*(\cdot)$ and random guessing. Random guessing is the base predictor with $cfb = 0.5$, as $H \perp\!\!\!\perp B$. For the predictor $h^*(\cdot)$, predictions take values from $\{h^*(0), h^*(1)\} = \{0.49, 0.54\}$. We anticipate that $cfb^* \geq 0.5$ as the performance of the random guessing should be no better than that of $h^*(\cdot)$. Note that we can use the probability triples to express the distribution of $(B \mid H^*)$ in this scenario. If we randomly select $\{(B_1, H_1), (B_2, H_2)\}$ from the population, the probability of getting concordant and discordant pairs is summarized in Table 1.

**Table 1** Joint probabilities of paired experiments

|               | $B_1 = B_2$ | $B_1 > B_2$ | $B_1 < B_2$ | Total |
| --- | --- | --- | --- | --- |
| $H_1^* = H_2^*$ | 0.28115 | 0.109425 | 0.109425 | 0.5 |
| $H_1^* > H_2^*$ | 0.135   | 0.05545  | 0.05955  | 0.25 |
| $H_1^* < H_2^*$ | 0.135   | 0.05955  | 0.05545  | 0.25 |
| Total         | 0.55115 | 0.224425 | 0.224425 | 1 |

The joint probability of two pairs matching on both $H^*$ and $B$ is calculated as $0.5^2(\sum_i p_i^2 + \sum_i q_i^2) = 0.28115$. Having a concordant pair, $(H_1^* > H_2^*, B_1 > B_2)$, consists of 3 mutually disjoint scenarios, and its probability is $0.5^2 \sum_{a>b} q_a p_b = 0.05545$, where $a$ and $b$ denote the level of $B$. Similarly, the probability of having a discordant pair, $(H_1^* < H_2^*, B_1 > B_2)$, is $0.5^2 \sum_{a<b} q_a p_b = 0.05955$. Thus, based on the definition (1), we obtain

$$cfb^* = \frac{0.05545 + 0.5 \cdot 0.109425}{0.224425} = 0.4908655,$$

which is smaller than 0.5. In other words, *cfb* fails to sensibly contrast the predictive performance of $h^*(\cdot)$, resulting in the best possible predictor having a *cfb* below that of random guessing, indicating that *cfb* is not a proper scoring rule.

The numerical example identified above is not the only one that yields $cfb^* < 0.5$. To generalize, $h^*(\cdot)$ produces a prediction $H^* \in \{h^*(0), h^*(1)\} = \{p_{+1} - p_{-1}, q_{+1} - q_{-1}\}$. We assume, without loss of generality, that $\Pr(X = 1) = 0.5$, and that $h^*(1) > h^*(0)$. Noting the relationships $p_0 = 1 - p_{+1} - p_{-1}$ and $q_0 = 1 - q_{+1} - q_{-1}$, we will thus have

$$q_{+1} - q_{-1} > p_{+1} - p_{-1}. \tag{2}$$

If the probability of having a concordant pair is smaller than that of having a discordant pair, $cfb^*$ will be smaller than 0.5. This happens whenever the triples satisfy the inequality $\sum_{a>b} q_a p_b < \sum_{a<b} q_a p_b$, which is equal to

$$q_{+1} - q_{-1} + q_{-1} p_{+1} < p_{+1} - p_{-1} + q_{+1} p_{-1}. \tag{3}$$

Thus, when both (2) and (3) hold simultaneously, $cfb^* < 0.5$. Figure 1 displays the $cfb^*$ values calculated from a large suite of $\{(p_{-1}, p_0, p_{+1}), (q_{-1}, q_0, q_{+1})\}$ values, obtained via a brute force simulation. The simulation searches combinations of probability triples that satisfy (2) and (3) on the interval [0, 1] (detailed simulation and calculation process are provided in Appendix A). These values of $cfb^*$ range from 0.4188 to 0.5 with a median of 0.4916. The smallest $cfb^*$ value, 0.4188, occurs when $(p_{-1}, p_0, p_{+1}) = (0.03, 0, 0.97)$ and $(q_{-1}, q_0, q_{+1}) = (0, 0.06, 0.94)$.

These improper scenarios can easily extend to a continuous $X$ (or a continuous $H^*$). One choice is through the connection between Bernoulli and Beta distributions. For instance, a binary $X$ with $\Pr(X = 1) = 0.5$ can be treated as the limit of a continuous distribution $Beta(\varepsilon, \varepsilon)$ as $\varepsilon$ approaches 0. Therefore, we consider a continuous $X$ having a Beta distribution and define $\Pr(B \mid X)$ as a linear interpolation of two sets of probabilities $(p_{-1}, p_0, p_{+1})$ and $(q_{-1}, q_0, q_{+1})$, which satisfy (2) and (3). That is, $\Pr(B = i \mid X) = p_i + (q_i - p_i)X$. It is mathematically guaranteed that by making parameters for the Beta distribution small enough, *cfb*\* < 0.5 (as demonstrated for the binary $X$). Appendix B provides two such examples that yield *cfb*\* < 0.5.

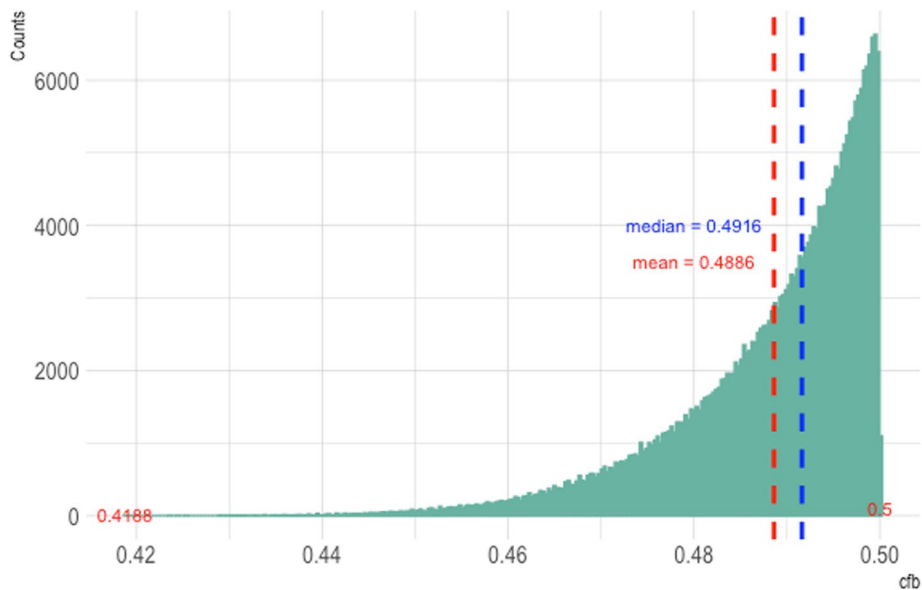

**Fig. 1** Histogram of *cfb*\* values calculated from the triples satisfying (2) and (3)

Note that in the context of binary risk prediction and the absence of censoring, the c-statistic (equal to the area under the receiver operating characteristic curve) is a proper scoring rule [17]. Why is *cfb* an improper scoring rule despite its conceptual analogy with the c-statistic? It is because there exist pairs of probability triples that satisfy both (2) and (3) simultaneously for a ternary $B$. Conversely, no such probability sets exist when $B$ is binary. In the case of c-statistic where the outcome is binary, (2) and (3) cannot hold simultaneously when we additionally require that $q_{-1} + q_{+1} = p_{-1} + p_{+1} = 1$. This finding raises concerns about applying the c-statistic to non-binary outcomes. Similar rank-based measures for other metrics that pertain to non-binary outcomes are also at risk of being improper. For instance, Blanche et al. (2019) demonstrated that the c-statistic is not a proper scoring rule for the concordance of time-to-event values [18].

*Improperness under the counterfactual framework*
Based on probability triples governing the distribution of $B$, we have characterized distributions of $(B \mid X)$ that yield *cfb*\* < 0.5. In counterfactual scenarios, the outcome that would be observed under treatment $T = t$ is denoted as $Y^{(t)}$. The population defined by $(Y^{(0)}, Y^{(1)} \mid X)$ *imposes* a distribution of $B = Y^{(1)} - Y^{(0)}$ given $X$, but we cannot necessarily recover an *arbitrary* pair of probability triples describing $(B \mid X)$ this way. However, probability distributions for $(B \mid X)$ derived from a distribution for $(Y^{(0)}, Y^{(1)} \mid X)$ can still result in *cfb*\* < 0.5. A way to specify such a population is by connecting counterfactual outcomes with the previously identified $(B \mid X)$ distributions that result in *cfb*\* < 0.5. Under the assumption $Y^{(0)} \perp\!\!\!\perp Y^{(1)} \mid X$, we can numerically evaluate whether or not there exist distributions of $(Y^{(0)} \mid X)$ and $(Y^{(1)} \mid X)$ that can produce a given $(B \mid X)$ distribution. We found a subset of identified probability triples that can arise from a counterfactual starting point and yield *cfb*\* < 0.5. The detailed screening process is provided in Appendix C. Similar to us, Hoogland et al. also cast the development of *cfb* in the counterfactual outcome framework as a starting point for their investigations [11].

***cfb* is sensitive to correlation between counterfactual outcomes**
Specifying the population distribution via counterfactuals reveals another caveat of *cfb*. We find that *cfb*\* changes as a function of the conditional dependence between counterfactual outcomes. This finding is more general, but it is more easily understood in the context of continuous counterfactual outcomes that have linear relationships with normally distributed baseline characteristics. The setting allows a closed-form expression for *cfb**, and the conditional dependency level is quantified by the correlation coefficient between unobserved random terms added to the linear relationships (see Appendix D for an example). We denote the correlation coefficient as $\rho$. In practice, as $Y^{(0)}$ and $Y^{(1)}$ cannot be observed simultaneously for the same individual, the conditional dependence is unidentifiable. But the fact is that $\rho$ impacts the distribution of $B$ and thus the value of *cfb*. The expression for dichotomous outcomes obtained by thresholding the continuous $Y^{(0)}$ and $Y^{(1)}$ does not lead to an obvious closed-form expression, but the dependence of *cfb* on $\rho$ stands.

### *cfb* is sensitive to definition of matched pairs

We consider a source population, specifically, its marginal distribution of $X$ and its conditional distribution of $(Y \mid T, X)$. In this context, to calculate *cfb* for a given $h(\cdot)$, a connection needs to be built between $(Y \mid T, X)$ and $(B \mid H)$. One way, as proposed by van Klaveren et al., is constructing a matched population consisting of matched patient pairs [5]. Specifically, in their original work, they used examples for $h(\cdot)$ based on a logistic regression model for the observed $Y$ given treatment and covariates and defined the observed benefit as the difference in outcomes between two similar patients in a matched pair. These two patients were from different treatment groups, and the similarity was defined in two ways: similarity in covariate patterns or similarity in predicted benefits. While the authors considered both definitions of matching acceptable, we show that matched pairs based on these two criteria can yield varying *cfb*s. Within each matched pair, we denote the quantities of interest as $\{(Y_1, X_1, T_1), (Y_2, X_2, T_2)\}$.

To give concrete examples, say we have an RCT with $T \perp\!\!\!\perp X$. The covariate $X$ is presumed to be a ternary variable taking value $x \in \{0, 1, 2\}$, and we set the distribution of $X$ as

$$\Pr(X = x) = \begin{cases} a, & x = 0 \\ b, & x = 1 \\ 1 - a - b, & x = 2. \end{cases}$$

It is convenient to parameterize $\Pr(Y = 1 \mid T, X)$ as

$$\text{logit}(E[Y \mid T, X]) = \beta_0 + \beta_x X + \beta_t T + \beta_{xt} TX.$$

In this setting, parameters $\{a, b, \beta_0, \beta_x, \beta_t, \beta_{xt}\}$ determines a distribution of $X$ and a distribution of $(Y \mid T, X)$, which are the basic ingredients for constructing a matched population. For the same ingredients, matching patients by $X$ or matching patients by $H$ can lead to different distributions of $(B \mid H)$ when $h(\cdot)$ is not a bijective function mapping $X$ to $H$ (see the Appendix E.2 for a mathematical derivation). Therefore, it is possible to generate substantially different *cfb* values from two distinct distributions of $(B \mid H)$. Such sensitivity to the definition of matched pairs is also highlighted by Hoogland et al. and investigated in more detail [11].

To illustrate the difference in *cfb* for the two definitions of matched pairs, we evaluate the discriminative ability of treatment benefit predictor $h(x) = x^2 - x - 1$ with prediction $H \in \{-1, 1\}$. This benefit predictor will be applied to different source populations, which are determined as follows. Parameters $a$ and $b$ take values from the sequence that starts at 0 with increment 0.01. Intending to screen as many combinations of $a$ and $b$ as possible, we consider all that satisfy $0 < a + b < 1$ and obtain 498,501 combinations. For each combination, $\beta_0, \beta_x, \beta_t$ and $\beta_{xt}$ are each generated independently from the uniform distribution on the interval $(-5, 5)$.

With the same set of parameters, we obtain two distributions of $(B \mid H)$ for matching on $X$ versus on $H$. We then compare the corresponding *cfb* calculated from the closed-form expression shown in Appendix A. Figure 2 shows the difference between the two *cfb*s among all generated source populations. In most cases, the absolute differences were smaller than 0.05. However, note that in some cases, the difference could be as much as 0.2466. This simple setup demonstrates that different definitions of matched pairs can induce substantial differences in *cfb* for the same $h(\cdot)$ in the same source population. When dealing with multiple covariates and a complex function $h(\cdot)$, the choice of matched pairs definition can have a considerable impact. Furthermore, when continuous predictors are present, exact matching will need to be relaxed to some form of close matching. Limited computational resources may further affect the choice of matching definition, potentially leading to less efficient results.

Further, even when consistently using the same definition, say matching by $X$, *cfb* is also sensitive to the sampling scheme for constructing matched pairs at the population level. One method of forming a matched population is to draw two independent patients from a joint distribution of $(Y, T, X)$ with conditioning on $X_1 = X_2$ and $T_1 = 1$ and $T_2 = 0$. However, this procedure can change the distribution of the covariate $X$ from that in the source population, which may not be desirable for inference purposes. Alternatively, we can first draw a treated patient from the distribution of $(Y, X \mid T = 1)$ and then sequentially select a control group patient with the same $X$. Under an RCT setting and with an infinite sample, the second procedure does not alter the distribution of $X$ (see Appendix E.1 for a detailed explanation). Thus, it is necessary to carefully define the sampling scheme for forming matched patient pairs to prevent altering the distribution of the covariate *X* in the source population.

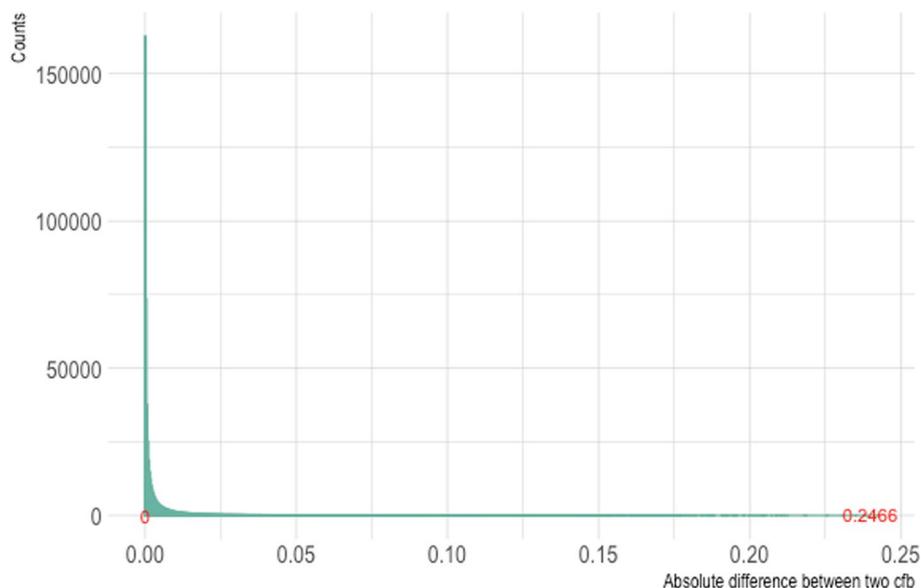

**Fig. 2** Histogram of the absolute difference between two *cfb* for matching factors *X* and *H*

## Discussion

In this work, we presented three fundamental problems of *cfb* through examples and theoretical developments. First, we showed that *cfb* is not a proper scoring rule. In particular, we found that the best possible predictor $h^*(x) = E[B \mid X = x]$ can result in a *cfb* that is lower than the *cfb* of a useless predictor based on chance prediction. Improperness is a grave concern as it can lead to misleading conclusions about the performance of predictors of treatment benefit.

Further, we showed that *cfb* is sensitive to the unestimable correlation between counterfactual outcomes conditional on covariates and is also sensitive to the definition of matched pairs. These issues are indeed interrelated. Under the counterfactual framework and RCT settings, van Klaveren et al. suggested using matched patient pairs to create the target distribution of $(B \mid H)$ from the distribution $(Y \mid T, X)$ [5]. But this matched population generally contains no information on the conditional dependency. The original work on *cfb* acknowledges this and makes it clear that the counterfactual outcomes under two treatment arms are assumed to be independent conditional on covariates [19]. Clearly, such a strong assumption could be violated in many applications, resulting in a *cfb* that might not be faithful to reality.

To keep the arguments intuitive, we demonstrated the aforementioned limitations of *cfb* in the context of a single explanatory variable. Involving multiple covariates will not alleviate the problems explained above. Ultimately, multivariate benefit predictors generate a scalar predicted benefit which can be considered a single explanatory variable. On the other hand, multivariate benefit predictors pose additional challenges, particularly with regard to matching. In addition to the sensitivity of *cfb* to the definition of matched pairs, the ultimate approximation required (e.g., specifying a maximum acceptable distance between matched high-dimensional *X*) can further undermine the accuracy of *cfb*.

For evaluating the discriminatory performance of benefit predictors, there are alternative metrics that are not suffering from these issues identified in this work. One particular metric is the Concentration of Benefit ($C_b$) [6]. $C_b$ is directly related to the Gini index and is concerned about the dispersion of the distribution of $E(B \mid X)$. It differs from a rank-based metric like *cfb* in that it remains a proper scoring rule, is not sensitive to unobserved correlation among counterfactuals, and does not require matching for its estimation. As such, as long as there are no ties in predicted benefits, a sample will lead to an unequivocal value of $C_b$. On a broader scale, how to develop and validate models for treatment benefit prediction is a nascent area of research, and critically evaluating the theoretical foundations and empirical performance of existing metrics should parallel the quest for the development of new ones.

## Appendix A: Screening improper scenarios with a binary *H**

This section provides the details of generating improper scenarios when $X$ is binary, and $\Pr(X=1) = c$. Suppose $H \in \{h_1, h_2\}$, $h_1 < h_2$ and $\Pr(H = h_2) = c$. We express $\Pr(B = i \mid H = h_1) = p_i$ and $\Pr(B = i \mid H = h_2) = q_i$, where $i = -1, 0, 1$. According to Eq. (1), the *cfb* is calculated as follows

$$\begin{aligned} cfb = & \frac{c \cdot (1-c)}{A}(q_{+1}p_0 + q_{+1}p_{-1} + q_0 p_{-1}) \\ & + \frac{0.5c^2}{A}(q_{+1}q_0 + q_{+1}q_{-1} + q_0 q_{-1}) \qquad (4) \\ & + \frac{0.5(1-c)^2}{A}(p_{+1}p_0 + p_{+1}p_{-1} + p_0 p_{-1}), \end{aligned}$$

where $A = c \cdot (1-c)(p_{+1}q_0 + p_{+1}q_{-1} + p_0 q_{-1} + q_{+1}p_0 + q_{+1}p_{-1} + q_0 p_{-1}) + c^2(q_{+1}q_0 + q_{+1}q_{-1} + q_0 q_{-1}) + (1-c)^2 (p_{+1}p_0 + p_{+1}p_{-1} + p_0 p_{-1})$.

The decimal place of the triples influences the number of important scenarios generated, and keeping more decimal places allows for having more important scenarios. The values of *cfb* in Fig. 1 are calculated based on the triples rounded to two decimal places, where $c = 0.5$. We generate 101 possible values from $[0, 1]$ for each probability in the probability triples. In total, we consider $101^4$ possible combinations of $(p_{-1}, p_0, p_{+1})$ and $(q_{-1}, q_0, q_{+1})$, and we search the distributions of $(B \mid X)$ that yield $cfb^* < 0.5$ from the $101^4$ possible combinations. We found 283,523 pairs of triples that give $cfb^* < 0.5$.

## Appendix B: Improper scenarios with a continuous *H**

For a continuous $X$, we provide two specific examples with $X \sim Beta(0.5, 0.5)$. Figure 3 depicts the density plot of $Beta(0.5, 0.5)$ with about probability 0.25 assigned to the interval $x \in (0.85, 1]$ and about probability 0.25 assigned to the interval $x \in [0, 0.15)$. The first improper scenario considers linear interpolations of the identified triples $(p_{-1}, p_0, p_{+1}) = (0.08, 0, 0.92)$ and $(q_{-1}, q_0, q_{+1}) = (0, 0.15, 0.85)$. The distribution of $(B \mid X)$ is shown in Fig. 4, where the green line represents $h^*(X)$. As there is no closed-form expression of *cfb* with a continuous $X$, the value of $cfb^*$ is approximated by numerical methods. With a Monte Carlo sample with size $10^7$, the value of $cfb^*$ is 0.4442 with a Monte Carlo standard error 0.0002.

The second improper scenario describes a less extreme situation, where the two triples involve no zero probabilities. The selected two triples are $(p_{-1}, p_0, p_{+1}) = (0.54, 0.37, 0.09)$ and $(q_{-1}, q_0, q_{+1}) = (0.68, 0.01, 0.31)$. Figure 5 illustrates the $\Pr(B = b \mid X = x)$, and the corresponding $h^*(X)$ is shown in green. Similarly, $cfb^* = 0.4906$ is calculated based on a Monte Carlo sample with $10^7$ observations, and the Monte Carlo standard error is 0.0002.

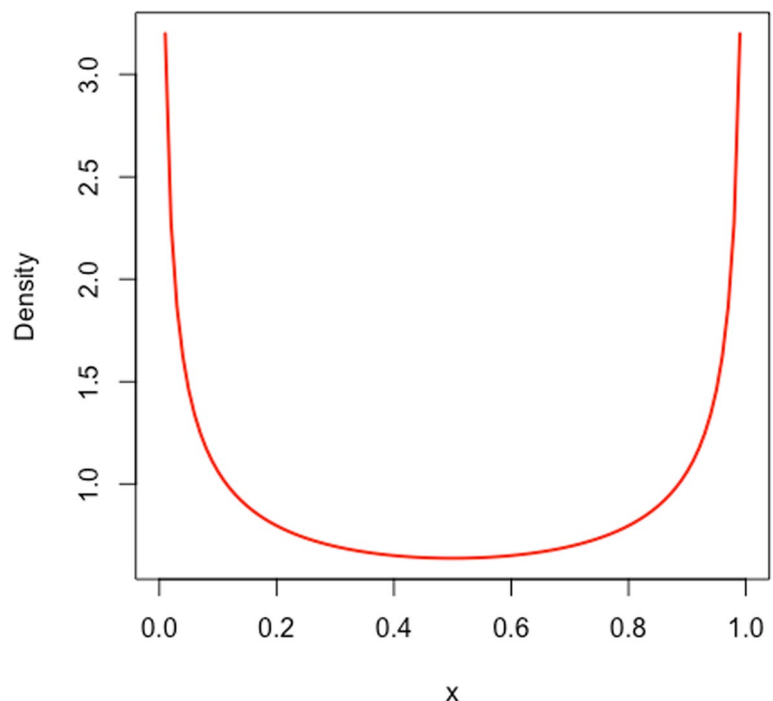

**Fig. 3** The density plot of $X \sim Beta(0.5, 0.5)$

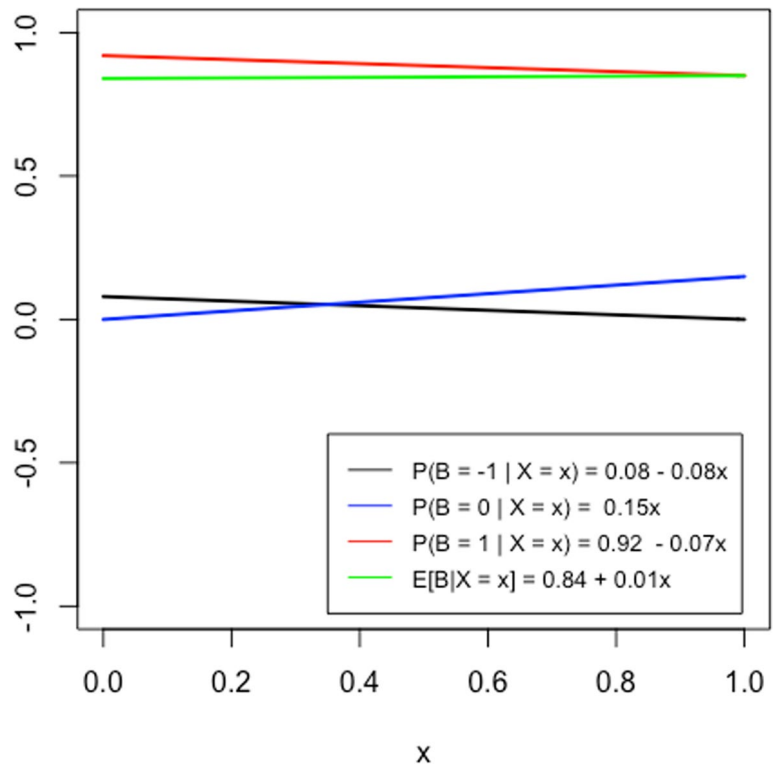

**Fig. 4** The probability $\Pr(B = b \mid X = x)$ and $h^*(X)$ for $(p_{-1}, p_0, p_{+1}) = (0.08, 0, 0.92)$ and $(q_{-1}, q_0, q_{+1}) = (0, 0.15, 0.85)$

## Appendix C: Screening improper scenarios under the counterfactual framework

We interpret $B$ in improper scenarios by $Y^{(0)}$ and $Y^{(1)}$. Suppose $X$ is binary and $h^*(X)$ is a bijection with $h^*(1) > h^*(0)$. We mathematically "screen" any given distribution of $(B \mid X)$ (e.g., pair of probability triples) for compatibility with some distribution of $(Y^{(0)}, Y^{(1)} \mid X)$.

Recall that we summarized the distribution by two probability triples: $(p_{-1}, p_0, p_{+1})$ and $(q_{-1}, q_0, q_{+1})$. Specifically,

$$\Pr(B = -1 \mid X = 0) = p_{-1}, \ \Pr(B = -1 \mid X = 1) = q_{-1},$$
$$\Pr(B = 0 \mid X = 0) = p_0, \ \Pr(B = 0 \mid X = 1) = q_0,$$
$$\Pr(B = 1 \mid X = 0) = p_{+1}, \ \Pr(B = 1 \mid X = 1) = q_{+1}.$$

Assume $Y^{(0)} \perp\!\!\!\perp Y^{(1)} \mid X$. The connections between the two probability triples and distributions of $(Y^{(0)} \mid X)$ and $(Y^{(1)} \mid X)$ are:

$$p_{-1} = \Pr(Y^{(0)} = 1 \mid X = 0)(1 - \Pr(Y^{(1)} = 1 \mid X = 0)),$$
$$p_{+1} = \Pr(Y^{(1)} = 1 \mid X = 0)(1 - \Pr(Y^{(0)} = 1 \mid X = 0)),$$
$$q_{-1} = \Pr(Y^{(0)} = 1 \mid X = 1)(1 - \Pr(Y^{(1)} = 1 \mid X = 1)),$$
$$q_{+1} = \Pr(Y^{(1)} = 1 \mid X = 1)(1 - \Pr(Y^{(0)} = 1 \mid X = 1)).$$

These four equations lead us to quadratic functions of $\Pr(Y^{(0)} = 1 \mid X = 0)$ and $\Pr(Y^{(0)} = 1 \mid X = 1)$, which can be expressed as:

$$\Pr(Y^{(1)} = 1 \mid X = 0)^2 + (p_{-1} - 1 - p_{+1})\Pr(Y^{(1)} = 1 \mid X = 0) + p_{+1} = 0.$$
$$\Pr(Y^{(1)} = 1 \mid X = 1)^2 + (q_{-1} - 1 - q_{+1})\Pr(Y^{(1)} = 1 \mid X = 1) + q_{+1} = 0.$$

The two quadratic functions have solutions when the discriminant of the quadratic function is greater than or equal to 0. As long as probability triples that satisfy inequalities (2) and (3) also satisfy inequalities:

$$(p_{-1} - 1 - p_{+1})^2 - 4p_{+1} \geq 0,$$
$$(q_{-1} - 1 - q_{+1})^2 - 4q_{+1} \geq 0,$$

we can find distributions of $(Y^{(0)}, Y^{(1)} \mid X)$ result in $cfb^* < 0.5$. The solutions of the quadratic functions are:

$$\Pr(Y^{(1)} = 1 \mid X = 0) = \frac{(p_{+1} + 1 - p_{-1}) \pm \sqrt{(p_{-1} - 1 - p_{+1})^2 - 4p_{+1}}}{2},$$
$$\Pr(Y^{(1)} = 1 \mid X = 1) = \frac{p_{-1}}{1 - \Pr(Y^{(0)} = 1 \mid X = 0)},$$
$$\Pr(Y^{(0)} = 1 \mid X = 0) = \frac{(q_{+1} + 1 - q_{-1}) \pm \sqrt{(q_{-1} - 1 - q_{+1})^2 - 4q_{+1}}}{2},$$
$$\Pr(Y^{(0)} = 1 \mid X = 1) = \frac{q_{-1}}{1 - \Pr(Y^{(1)} = 1 \mid X = 0)}.$$

In other words, as long as the discrimination of a quadratic function is greater than or equal to 0, there exists at least one real number solution for the corresponding quadratic function. It implies that the distributions

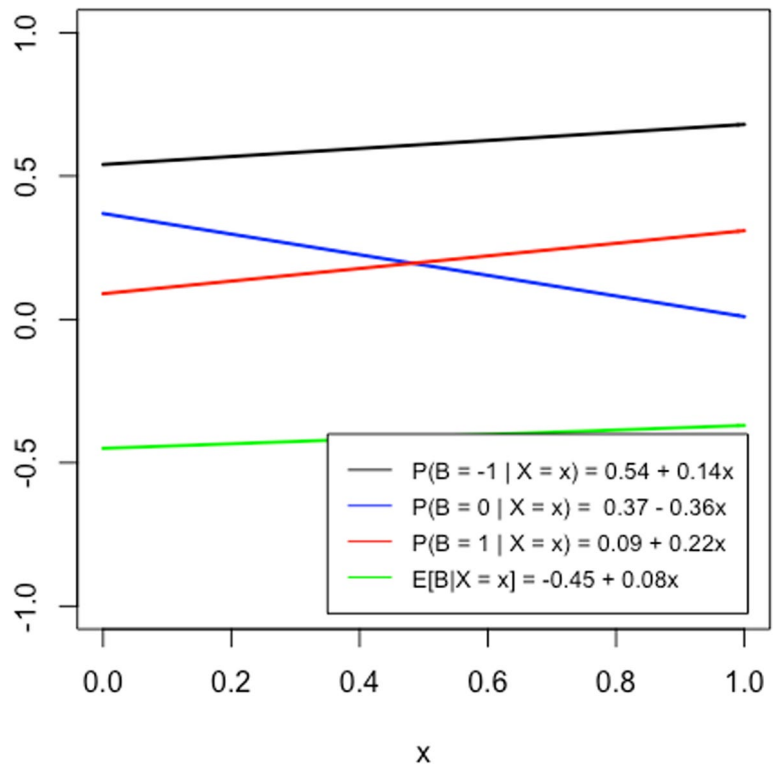

**Fig. 5** The probability $\Pr(B = b \mid X = x)$ and $h^*(X)$ for $(p_{-1}, p_0, p_{+1}) = (0.54, 0.37, 0.09)$ and $(q_{-1}, q_0, q_{+1}) = (0.68, 0.09, 0.31)$

of $(Y^{(0)}, Y^{(1)} \mid X)$ yield improper scenarios that result $cfb^* < 0.5$.

To illustrate the extension of examples in Fig. 1, we shaded the probability triples satisfied the four inequalities in black, which is displayed in Fig. 6. We find that only a subset of the previously found $(B \mid X)$ distributions with $cfb^* < 0.5$ can arise from a counterfactual starting point, with this subset having relatively larger $cfb$ values. Particularly, within the shaded area, the maximum $cfb$ is 0.5 and the minimum is 0.4830. The mean and median of $cfb$ are close to the maximum, which are 0.4961 and 0.4969 respectively. But the overriding point is that there are distributions of $(Y^{(0)}, Y^{(1)} \mid X)$ which yield $cfb^* < 0.5$.

One step further, if $\Pr(Y^{(0)} = 1 \mid X)$ and $\Pr(Y^{(1)} = 1 \mid X)$ are described by logistic regressions:

$$\text{logit}\left(\text{E}[Y^{(0)} \mid X]\right) = \beta_0 + \beta_x X,$$
$$\text{logit}\left(\text{E}[Y^{(1)} \mid X]\right) = \beta_0 + \beta_x X + \beta_t + \beta_{xt} X.$$

Each distribution of $(Y^{(0)}, Y^{(1)} \mid X)$ can be mapped to a set of logistic regression model parameters that give $cfb^* < 0.5$. For each pair of probability triples satisfying both (2) and (3), there exists a set of the logistic regression parameters, $\{\beta_0, \beta_x, \beta_t, \beta_{xt}\}$, that yields $cfb^* < 0.5$ for binary $X$. Specifically, the parameters are

$$\beta_0 = \text{logit}\left(\Pr(Y^{(0)} = 1 \mid X = 0)\right),$$
$$\beta_x = \text{logit}\left(\Pr(Y^{(0)} = 1 \mid X = 1)\right) - \text{logit}\left(\Pr(Y^{(0)} = 1 \mid X = 0)\right),$$
$$\beta_t = \text{logit}\left(\Pr(Y^{(1)} = 1 \mid X = 0)\right) - \text{logit}\left(\Pr(Y^{(0)} = 1 \mid X = 0)\right),$$
$$\beta_{xt} = \text{logit}\left(\Pr(Y^{(1)} = 1 \mid X = 1)\right) - \beta_t - 1 + \beta_0.$$

We have demonstrated that the improper scenarios involving a continuous $X$ can be constructed based on the scenarios with a binary $X$. The same reasoning and process can be used to find a distribution of $(Y^{(0)}, Y^{(1)}, X)$ yielding $cfb^* < 0.5$ with a continuous variable $X$.

## Appendix D: Correlation between counterfactual outcomes

Consider a continuous benefit $B$ with continuous variables $Y^{(0)}$ and $Y^{(1)}$. Suppose $X$ is generated from a standard normal distribution, and counterfactual outcomes are characterized by linear functions:

$$Y^{(0)} = \beta_0 + \beta_x X + \varepsilon_0,$$
$$Y^{(1)} = (\beta_0 + \beta_t) + (\beta_x + \beta_{xt})X + \varepsilon_1,$$

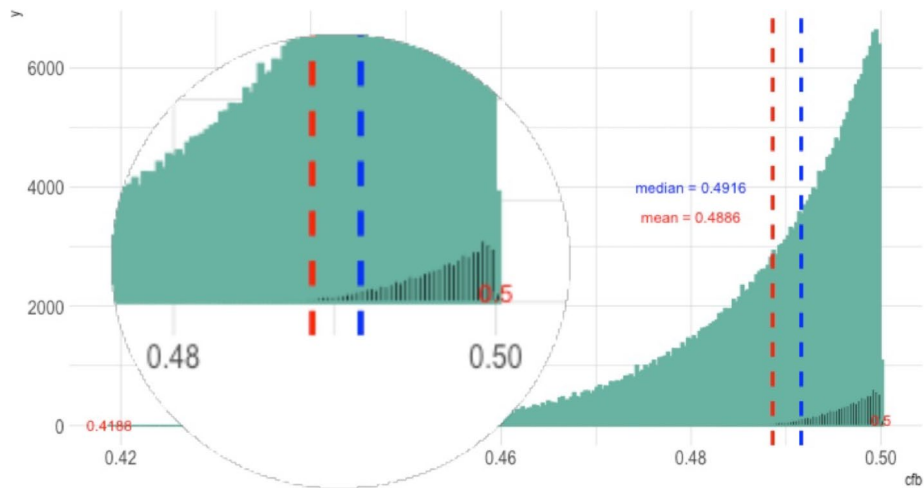

**Fig. 6** Extension of Fig. 1 (y-axis represents counts). The shaded part shows that distributions of $(Y^{(0)}, Y^{(1)} \mid X)$ yields $cfb^* < 0.5$

where $\varepsilon_0$ and $\varepsilon_1$ are potentially correlated random variables following the normal distribution with mean 0, variance $\sigma^2$, and $\varepsilon_0, \varepsilon_1 \perp\!\!\!\perp X$. In general, distributions of $\varepsilon_0$ and $\varepsilon_1$ could be different. The parameter $\beta_t$ is interpreted as the main treatment effect, and $\beta_{xt}$ is the interaction treatment effect.

We denote correlation coefficient of $\varepsilon_0$ and $\varepsilon_1$ as $\rho$. We obtain *cfb* for $h^*(X)$, which is $cfb^* = 2F_{H_2^* - H_1^*, B_2 - B_1}((0,0)^T; \mu, \Sigma)$, where $B \sim N(\beta_t, \beta_{xt}^2 + 2\sigma^2(1-\rho))$ and $H^* \sim N(\beta_t, \beta_{xt}^2)$. The mean and the covariance matrix are

$$\mu = \begin{pmatrix} 0 \\ 0 \end{pmatrix}, \ \Sigma = \begin{pmatrix} 2\beta_{xt}^2 & 2\beta_{xt}^2 \\ 2\beta_{xt}^2 & 2\beta_{xt}^2 + 4(1-\rho)\sigma^2 \end{pmatrix}. \quad (5)$$

Therefore, the unobserved correlation $\rho$ has an impact on the value of *cfb*.

## Appendix E: Definition of matched patient pairs to construct matched populations

Recall that the two ingredients for constructing a matched population are a distribution of $X$ and a distribution of $(Y \mid T, X)$. We assume that it is possible to obtain a distribution of $(B \mid H)$ from the two ingredients and a given benefit predictor $h(\cdot)$.

### Appendix E.1: Sampling schemes

Our analysis reveals that matching on the same variable $X$ using two different sampling schemes can produce matched populations that differ significantly. To illustrate this point, we first randomly and independently draw two patients from $(Y, T, X)$, i.e., $\{(Y_1, T_1, X_1), (Y_2, T_2, X_2)\}$, with conditioning on $X_1 = X_2$ and $T_1 = 1$ and $T_2 = 0$. We use $\Pr(X' = x)$ to represent the distribution of $X$ after matching. Then $\Pr(Y_1, Y_2, X' \mid T_1 = 1, T_2 = 0)$ can be expressed as:

$$\Pr(Y_1, Y_2, X' \mid T_1 = 1, T_2 = 0)$$
$$= \Pr(Y_1, Y_2 \mid T_1 = 1, T_2 = 0, X_1 = X_2 = x) \Pr(X_1 = X_2 = x)$$
$$= \Pr(Y_1 \mid T_1 = 1, X_1 = x) \Pr(Y_2 \mid T_2 = 0, X_2 = x) \frac{\Pr(X = x)^2}{\sum_x \Pr(X = x)^2},$$

where $\Pr(X' = x) = \frac{\Pr(X=x)^2}{\sum_x \Pr(X=x)^2}$. It shows that the distribution of $X$ changes after matching, which has a less peaked distribution than that of $X'$.

Second, the subjects are drawn sequentially. The matched population is constructed by first drawing a treated patient from $(Y, T, X)$, then drawing a control patient with the same $X$. In this case, the $\Pr(Y_1, Y_2, X' \mid T_1 = 1, T_2 = 0)$ can be expressed as:

$$\Pr(Y_1, Y_2, X' \mid T_1 = 1, T_2 = 0)$$
$$= \Pr(Y_1 \mid T_1 = 1, X_1 = x) \Pr(X_1 = x \mid T_1 = 1) \Pr(Y_2 \mid T_2 = 0, X_2 = x),$$

where $\Pr(X' = x) = \Pr(X = x \mid T = 1)$. The marginal distribution of $X'$ equals the distribution of $X$ among the treated patients in the source population. Under an RCT setting with $X \perp\!\!\!\perp T$, the second procedure provides $\Pr(X' = x) = \Pr(X = x)$.

### Appendix E.2: Matching factors

In [5], covariates $X$ and predicted treatment benefit $H$ were considered as matching factors. We assume that the sampling scheme applied does not change the original marginal distribution of $X$ under the RCT settings and constructs matched populations based on different matching factors $X$ and $H$.

Given a treatment benefit predictor $h(x)$, we show the connection between a distribution $(B \mid H)$ and a distribution $(Y \mid T, X)$ for matching factors $X$ and $H$, respectively. If a matched pair is matched by $X$, then the distribution of $(B \mid H)$ can be expressed as

$$\sum_{x \in \mathcal{S}} \Pr(B = b, X = x \mid H)$$
$$= \sum_{x \in \mathcal{S}} \Pr(B = b \mid X = x) \Pr(X = x \mid H)$$
$$= \sum_{x \in \mathcal{S}} \left( \sum_{(y_1, y_2) \in \mathcal{B}_X} \Pr(Y_1 = y_1 \mid T_1 = 1, X = x) \Pr(Y_2 = y_2 \mid T_2 = 0, X = x) \right) \times \Pr(X = x \mid H),$$

where set $\mathcal{S}$ contains all possible $x$ such that $h(x) = H, x \in X$, and set $\mathcal{B}_X$ consists of all possible $(y_1, y_2)$ pairs that satisfy $\Pr(B = b \mid X) = \Pr(Y_1 = y_1 \mid T_1 = 1, X) \Pr(Y_2 = y_2 \mid T_2 = 0, X)$.

If matched pairs are matched by $H$, it is possible for patients in a matched pair to have different values of the covariate $X$. Specifically, a matched population is created by first selecting a treated patient with $(Y_1 = y_1, T_1 = 1, X_1 = x_1)$ where $h(x_1) = H$. Then, another patient is selected with $(Y_2 = y_2, T_2 = 0, X_2 = x_2)$ where $h(x_2) = h(x_1) = H$. Thus, the distribution of $(B \mid H)$ can be expressed as

$$\sum_{x_1 \in \mathcal{S}} \sum_{x_2 \in \mathcal{S}} \Pr(B = b, X_1 = x_1, X_2 = x_2 \mid H)$$
$$= \sum_{x_1 \in \mathcal{S}} \sum_{x_2 \in \mathcal{S}} \Pr(B = b \mid X_1 = x_1, X_2 = x_2) \Pr(X_1 = x_1, X_2 = x_2 \mid H)$$
$$= \sum_{x_1 \in \mathcal{S}} \sum_{x_2 \in \mathcal{S}} \left( \sum_{(y_1, y_2) \in \mathcal{B}_H} \Pr(Y_1 = y_1 \mid T_1 = 1, X_1 = x_1) \Pr(Y_2 = y_2 \mid T_2 = 0, X_2 = x_2) \right)$$
$$\times \Pr(X_1 = x_1 \mid H) \Pr(X_2 = x_2 \mid H).$$

Similarly, set $\mathcal{B}_\mathcal{H}$ consists of all $(y_1, y_2)$ pairs that make the equivalence holds. When $h(\cdot)$ is an invertible function with domain $X$ and codomain $H$, matching on $X$ is equivalent to matching on $H$ as the two joint distributions referred to as the two matched populations are the same.

**Acknowledgements**
Not applicable.

**Authors' contributions**
YX led the derivations, programming, and writing of the paper under the supervision of PG and MS. All three authors approved the final manuscript.

**Funding**
The research was supported by NSERC Discovery Grant RGPIN-2019-03957.

**Availability of data and materials**
The datasets illustrating the benefit predictor evaluation processes in the study are available in the 2022cfb repository (https://github.com/LilyYuanXia/2022cfb).

**Declarations**

**Ethics approval and consent to participate**
Not applicable.

**Consent for publication**
Not applicable.

**Competing interests**
The authors declare that they have no competing interests.

Received: 28 November 2022   Accepted: 17 April 2023
Published online: 16 May 2023

**Publisher's Note**
Springer Nature remains neutral with regard to jurisdictional claims in published maps and institutional affiliations.